\documentclass[twocolumn,english,showpacs,superscriptaddress,nofootinbib,prc]{revtex4-1}
\usepackage[T1]{fontenc}
\usepackage[latin9]{luainputenc}
\usepackage{babel}
\usepackage{float}
\usepackage{booktabs}
\usepackage{amsmath}
\usepackage{amssymb}
\usepackage{graphicx}
\usepackage{color}
\usepackage[unicode=true,pdfusetitle,
 bookmarks=true,bookmarksnumbered=false,bookmarksopen=false,
 breaklinks=false,pdfborder={0 0 1},backref=false,colorlinks=false]
 {hyperref}

\makeatletter

\providecommand{\tabularnewline}{\\}

\makeatother

\begin{document}
\title{Production of light-flavor and single-charmed hadrons in $pp$ collisions at $\sqrt{s}=5.02$ TeV in an equal-velocity quark combination model}
\author{Hai-hong Li}
\affiliation{School of Physical Science and Intelligent Engineering, Jining University, Shandong 273155, China}
\author{Feng-lan Shao}
\email{shaofl@mail.sdu.edu.cn}

\affiliation{School of Physics and Physical Engineering, Qufu Normal University, Shandong 273165, China}
\author{Jun Song }
\email{songjun2011@jnxy.edu.cn}
\affiliation{School of Physical Science and Intelligent Engineering, Jining University, Shandong 273155, China}

\begin{abstract}
    We apply an equal-velocity quark combination model to study the production of light-flavor hadrons and single-charmed hadrons at midrapidity in $pp$ collisions at $\sqrt{s}=5.02$ TeV. We find experimental data for $p_{T}$ spectra of $\Omega$ and $\phi$ exhibit a quark number scaling property, which is a clear signal of quark combination mechanism at hadronization. Experimental data for $p_T$ spectra of $p$, $\Lambda$, $\Xi$, $\Omega$, $\phi$ and $K^{*0}$ are systematically described by the model. The non-monotonic $p_{T}$ dependence of $\Omega/\phi$ ratio is naturally explained and we find it is closely related to the shape of the logarithm of strange quark $p_{T}$ distribution. Using $p_{T}$ spectra of light-flavor quarks obtained from light-flavor hadrons and a $p_T$ spectrum of charm quarks which is consistent with perturbative QCD calculations, the experimental data for differential cross-sections of $D^{0,+}$, $D_{s}^{+}$ and $\Lambda_{c}^{+}$ as the function of $p_{T}$ are systematically described. We predict the differential cross-sections of $\Xi_{c}^{0,+}$ and $\Omega_{c}^{0}$.  The ratio $\Xi_{c}^{0,+}/D^{0}$ in our model is about 0.16 and $\Omega_{c}^{0}/D^{0}$ is about 0.012 due to the cascade suppression of strangeness. In addition, the predicted $\Xi_{c}^{0,+}/D^{0}$ and $\Omega_{c}^{0}/D^{0}$ ratios exhibit the non-monotonic dependence on $p_{T}$ in the low $p_{T}$ range. 
\end{abstract}
\maketitle

\section{Introduction\label{sec:Intro}}

Recently, experiments of $pp$ and $p$Pb collisions at energies available
at the CERN Large Hadron Collider (LHC) found a series of interesting
properties of hadron production such as ridge and collectivity \citep{Khachatryan:2010gv,CMS:2012qk,Khachatryan:2015waa,Khachatryan:2016txc},
enhancement of strangeness and ratios of baryon to meson \citep{ALICE:2017jyt,Adam:2015vsf,Adam:2016dau,Abelev:2013haa}.
These striking observations are possibly related to a hot topic in
strong interactions, i.e., the formation of a small droplet of Quark-Gluon
Plasma (QGP) in $pp$ and $p$Pb collisions. Theoretical studies along
this direction are extensively carried out in the last few years from
different aspects. The key point is how to understand and simulate
the small final-state parton system created in $pp$ and $p$Pb
collisions at LHC energies. These studies usually focus on the application
of hydrodynamics to simulate mini-QGP evolution \citep{Luzum:2009sb,Liu:2011dk,Werner:2010ss,Bzdak:2013zma,Bozek:2013uha,Prasad:2009bx,Avsar:2010rf,Zhao:2017rgg},
the search of new features in string formation just before hadronization by various mechanisms
such as color re-connection \citep{Bautista:2015kwa,Bierlich:2014xba,Ortiz:2013yxa,Christiansen:2015yqa,Bierlich:2015rha},
the search of new features in string(cluster) fragmentation or parton(quark)
coalescence mechanism at hadronization \citep{Pop:2012ug,Minissale:2020bif,Gieseke:2017clv,Song:2017gcz,He:2019tik,Song:2018tpv},
etc. 

In our recent studies on $pp$ collisions at two collision energies
$\sqrt{s}=7,13$ TeV \citep{Song:2018tpv,Gou:2017foe,Zhang:2018vyr}
and on $p$Pb collisions at $\sqrt{s_{NN}}=5.02$ TeV \citep{Song:2017gcz,Li:2017zuj},
we found that an equal-velocity combination mechanism of constituent
quarks and antiquarks at hadronization can systematically describe
the experimental data for $p_{T}$ spectra of light-flavor hadrons
and single-charmed hadrons in the low and intermediate $p_{T}$ range
in these collision systems. The constituent quark degrees of
freedom just before hadronization play an important role for hadron
production in these collisions, which may be related to possible formation
of QGP droplet in $pp$ and $p$Pb collisions at LHC energies. Compared
with the traditional fragmentation mechanism usually applied in small
collision systems, this quark-combination ``new'' feature at hadronization
should be studied further with help of experimental data in $pp$
and $p$Pb collisions at other collision energies at LHC. 

In this paper, we use an equal-velocity quark combination model to
study the production of light-flavor hadrons and single-charmed hadrons
in $pp$ collisions at $\sqrt{s}=5.02$ TeV. Firstly, we use the model
to describe the experimental data of light-flavor hadrons \citep{Tripathy:2018ehz,Acharya:2019yoi,Sirunyan:2019rfz}.
We pay particular attention to how to systematically relate the observed
properties of hadrons to quark $p_T$ spectra at hadronization. For example, we correlate $p_{T}$ spectrum of $\Omega$ and that of $\phi$ by a scaling method to directly relate to $p_{T}$ distribution
of strange quarks at hadronization. Another example is that we can
relate the non-monotonic $p_{T}$ dependence of $\Omega/\phi$ ratio
to the shape of the logarithm of strange quark distribution.
Second, using $p_{T}$ spectra of light-flavor quarks obtained from
study of light-flavor hadrons and a charm quark distribution which
is consistent with perturbative QCD calculations, we further study
the equal-velocity combination of light-flavor and charm (anti-)quarks to explain the production
properties of single-charmed hadrons. We compare model results with
experimental data for differential cross-sections of $D^{0,+}$, $D_{s}^{+}$
and $\Lambda_{c}^{+}$ \citep{Acharya:2020lrg,Acharya:2019mgn}. We
predict the differential cross-sections of $\Xi_{c}^{0,+}$ and $\Omega_{c}^{0}$
and several baryon to meson ratios such as $\Xi_{c}^{0,+}/D^{0}$
and $\Omega_{c}^{0}/D^{0}$ for the future test. 

The paper is organized as follows. In Sec. \ref{sec:sdqcm}, we briefly
introduce a particular quark combination model under equal-velocity
combination approximation. In Sec. \ref{sec:LightFlavor_Results},
we show results for $p_{T}$ spectra of light-flavor hadrons in $pp$
collisions at $\sqrt{s}=5.02$ TeV. In Sec. \ref{sec:HF_results},
we show results for $p_{T}$ spectra and spectrum ratios of single-charmed
hadrons. In Sec. \ref{sec:summary}, we give the summary. 

\section{A brief introduction of equal-velocity quark combination model \label{sec:sdqcm}}

In this section, we briefly introduce a particular quark combination
model proposed in recent work \citep{Song:2017gcz}. This model applies
a simplified combination criterion, i.e., the equal-velocity combination
(EVC), to determine how constituent quarks and antiquarks at hadronization
form hadrons. This EVC model was inspired by the quark number scaling
property found in $p_{T}$ spectra of strange hadrons \citep{Song:2017gcz,Zhang:2018vyr}.
The model has successfully described $p_{T}$ spectra of light-flavor
hadrons and single-charmed hadrons in ground state in $pp$ collisions
at $\sqrt{s}=7,13$ TeV and in $p$Pb collisions at $\sqrt{s_{NN}}=5.02$
 TeV. Our latest studies on elliptic flow and $p_{T}$ spectra of hadrons
in relativistic heavy-ion collisions \citep{Song:2019sez,Song:2021ygg,Song:2020lpr,Wang:2019fcg}
also support the EVC model . 

In the scenario of stochastic combination of quarks and antiquarks
at hadronization, momentum distribution of the formed hadron $f_{h}\left(p\right)\equiv dN_{h}/dp$
can be constructed by those of quarks and antiquarks, 

\begin{align}
f_{B_{j}}(p_{B}) & =\int\mathrm{d}p_{1}\mathrm{d}p_{2}\mathrm{d}p_{3}{\cal R}_{B_{j}}(p_{1},p_{2},p_{3};p_{B})\,f_{q_{1}q_{2}q_{3}}(p_{1},p_{2},p_{3}),\label{eq:fb}\\
f_{M_{j}}(p_{M}) & =\int\mathrm{d}p_{1}\mathrm{d}p_{2}{\cal R}_{M_{j}}(p_{1},p_{2};p_{M})f_{q_{1}\bar{q}_{2}}(p_{1},p_{2}),\label{eq:fm}
\end{align}
where $f_{q_{1}q_{2}q_{3}}(p_{1},p_{2},p_{3})$ and $f_{q_{1}\bar{q}_{2}}(p_{1},p_{2})$
are joint momentum distributions for $q_{1}q_{2}q_{3}$ and $q_{1}\bar{q}_{2}$,
respectively. ${\cal R}_{B_{j}}(p_{1},p_{2},p_{3};p_{B})$ is the
combination probability function for three quarks $q_{1}q_{2}q_{3}$
with momenta $p_{1}$, $p_{2}$ and $p_{3}$ forming a baryon $B_{j}$
with quark composition $q_{1}q_{2}q_{3}$ and momentum $p_{B}$. ${\cal R}_{M_{j}}(p_{1},p_{2};p_{M})$
has similar meaning. 

Under the approximation of EVC, a hadron is formed by the combination of constituent
quarks and/or antiquarks with same velocity. Because momentum has
property $p_{i}=\gamma m_{i}v\propto m_{i}$ at the given velocity,
the momentum of the participant (anti-)quark $p_{i}$ should has a
particular fraction $x_{i}=p_{i}/p\propto m_{i}$ of the momentum of hadron $p$ where $m_{i}$ is the constituent mass of the quark $i$.
Considering the momentum conservation $\sum_{i}p_{i}=p$ we obtain 
\begin{equation}
x_{i}=\begin{cases}
\frac{m_{i}}{m_{1}+m_{2}+m_{3}} & i=1,2,3\,\,\,\textrm{for}\,B(q_{1}q_{2}q_{3})\\
\frac{m_{i}}{m_{1}+m_{2}} & i=1,2\,\,\,\,\,\,\,\,\textrm{for}\,M(q_{1}\bar{q}_{2})
\end{cases}.
\end{equation}
The constituent masses of quarks are taken as $m_{u}=m_{d}=0.3$ GeV,
$m_{s}=0.5$ GeV and $m_{c}=1.5$ GeV. The combination function therefore
has the simple form 
\begin{align}
{\cal R}_{B_{j}}(p_{1},p_{2},p_{3};p_{B}) & =\kappa_{B_{j}}\prod_{i=1}^{3}\delta(p_{i}-x_{i}p_{B}),\label{eq:RB}\\
{\cal R}_{M_{j}}(p_{1},p_{2};p_{M}) & =\kappa_{M_{j}}\prod_{i=1}^{2}\delta(p_{i}-x_{i}p_{M}).\label{eq:RM}
\end{align}
 $\kappa_{B_{j}}$ and $\kappa_{M_{j}}$ are independent of momentum
but can be dependent on numbers of (anti-) quarks at hadronization
and the property of the formed hadron such as spin. 

Substituting combination functions Eqs. (\ref{eq:RB}) and (\ref{eq:RM})
into Eqs. (\ref{eq:fb}) and (\ref{eq:fm}), we obtain
\begin{align}
f_{B_{j}}\left(p_{B}\right) & =\kappa_{B_{j}}\,f_{q_{1}q_{2}q_{3}}\left(x_{1}p_{B},x_{2}p_{B},x_{3}p_{B}\right),\label{eq:fbi}\\
f_{M_{j}}\left(p_{M}\right) & =\kappa_{M_{j}}f_{q_{1}\bar{q}_{2}}\left(x_{1}p_{M},x_{2}p_{M}\right).\label{eq:fmi}
\end{align}
Integrating above equations over the momentum, we obtain the number
of the formed hadrons
\begin{align}
N_{B_{j}} & =\kappa_{B_{j}}\int{\rm d}p_{B}f_{q_{1}q_{2}q_{3}}\left(x_{1}p_{B},x_{2}p_{B},x_{3}p_{B}\right),\label{eq:NBi_ini}\\
N_{M_{j}} & =\kappa_{M_{j}}\int{\rm d}p_{M}f_{q_{1}\bar{q}_{2}}\left(x_{1}p_{M},x_{2}p_{M}\right).\label{eq:NMi_ini}
\end{align}

The integral of joint momentum distribution of (anti-)quarks can be
rewritten as 
\begin{align}
\int{\rm d}p_{B}f_{q_{1}q_{2}q_{3}}\left(x_{1}p_{B},x_{2}p_{B},x_{3}p_{B}\right) & =\frac{N_{q_{1}q_{2}q_{3}}}{A_{B_{j}}},\label{eq:coe_ABi}\\
\int{\rm \mathrm{d}}p_{M}f_{q_{1}\bar{q}_{2}}\left(x_{1}p_{M},x_{2}p_{M}\right) & =\frac{N_{q_{1}\bar{q}_{2}}}{A_{M_{j}}}\label{eq:coe_AMi}
\end{align}
with 
\begin{align}
    N_{q_{1}q_{2}q_{3}} & =\iiint\mathrm{d}p_{1}\mathrm{d}p_{2}\mathrm{d}p_{3}f_{q_{1}q_{2}q_{3}}\left(p_{1,}p_{2},p_{3}\right),  \label{Nq1q2q3}\\
    N_{q_{1}\bar{q}_{2}} & =\iint\mathrm{d}p_{1}\mathrm{d}p_{2}f_{q_{1}\bar{q}_{2}}\left(p_{1},p_{2}\right). \label{Nq1q2bar}
\end{align}
Here, $N_{q_{1}\bar{q}_{2}}$ is the number of all $q_{1}\bar{q}_{2}$
pairs at hadronization. In general, we have $N_{q_{1}\bar{q}_{2}}=N_{q_{1}}N_{\bar{q}_{2}}$
where $N_{q_{1}}$ is the number of $q_{1}$ in system and $N_{\bar{q}_{2}}$
is that of $\bar{q}_{2}$. $N_{q_{1}q_{2}q_{3}}$ is the number of
all possible $q_{1}q_{2}q_{3}$ combinations. In general, $N_{q_{1}q_{2}q_{3}}$
equals to $N_{q_{1}}N_{q_{2}}N_{q_{3}}$ for different quark flavors,
$N_{q_{1}}(N_{q_{1}}-1)N_{q_{2}}$ for two identical quark flavor
and $N_{q_{1}}(N_{q_{1}}-1)(N_{q_{1}}-2)$ for three identical quark
flavor. Coefficients $A_{B_{j}}$ and $A_{M_{j}}$ are thus introduced
to characterize the effect of joint momentum distribution of (anti-)quarks
with correlated momenta on the number of the formed hadron. 

Substituting Eqs. (\ref{eq:coe_ABi}) and (\ref{eq:coe_AMi}) into
Eqs. (\ref{eq:NBi_ini}) and (\ref{eq:NMi_ini}), we obtain 

\begin{align}
N_{B_{j}} & =N_{q_{1}q_{2}q_{3}}\frac{\kappa_{B_{j}}}{A_{B_{j}}}=N_{q_{1}q_{2}q_{3}}P_{q_{1}q_{2}q_{3}\rightarrow B_{j}},\label{eq:NBi_gen}\\
N_{M_{j}} & =N_{q_{1}\bar{q}_{2}}\frac{\kappa_{M_{j}}}{A_{M_{j}}}=N_{q_{1}\bar{q}_{2}}P_{q_{1}\bar{q}_{2}\rightarrow M_{j}}.\label{eq:NMi_gen}
\end{align}
Coefficient ratio $\kappa_{B_{i}}/A_{B_{i}}$ thus has an intuitive
physical meaning, that is, the momentum-integrated probability of
$q_{1}q_{2}q_{3}$ forming a $B_{j}$. Therefore, we denote it as
$P_{q_{1}q_{2}q_{3}\rightarrow B_{j}}$ in the second equality. $\kappa_{M_{j}}/A_{M_{j}}$
denotes the momentum-integrated probability of a $q_{1}\bar{q}_{2}$
pair forming a $M_{j}$, and we denote it as $P_{q_{1}\bar{q}_{2}\rightarrow M_{j}}$. 

Because of the non-perturbative nature of $P_{q_{1}q_{2}q_{3}\rightarrow B_{i}}$
and $P_{q_{1}\bar{q}_{2}\rightarrow M_{i}}$, we will parameterize
them in the following text. Here, we consider the formation of hadrons
in two sectors. One is light-flavor hadrons which are exclusively
composed of light-flavor (anti-)quarks. Another is single-charmed
hadrons which are composed of a charm (anti-)quark and light-flavor
(anti-)quark(s). For convenience, light-flavor quarks are denoted
as $l_{i}$ ($l_{i}=d,u,s$) and $N_{l_{i}}$ their numbers. The number
of all light-flavor quarks is $N_{l}=\sum_{l_{i}}N_{l_{i}}$ and similar
for anti-quarks. Charm quarks are denoted as $c$ and $N_{c}$ its
number. 

Considering the stochastic feature of the quark combination and flavor
independence of strong interaction, the combination probability of
light-flavor (anti-)quarks can be parameterized by
\begin{eqnarray}
P_{l_{1}l_{2}l_{3}\rightarrow B_{j}} & = & C_{B_{j}}N_{iter}\frac{\overline{N}_{B}}{N_{lll}},\label{prob_B_lf}\\
P_{l_{1}\bar{l}_{2}\rightarrow M_{j}} & = & C_{M_{j}}\frac{\overline{N}_{M}}{N_{l\bar{l}}},\label{prob_M_lf}
\end{eqnarray}
where we use $\overline{N}_{B}/N_{lll}$ to denote the average probability
of three light-flavor quarks combining into a baryon and $\overline{N}_{M}/N_{l\bar{l}}$
to denote the average probability of a light-flavor quark and antiquark
pair combining into a meson. Here, $\overline{N}_{B}$ and $\overline{N}_{M}$
are the average number of all light-flavor baryons and that of all
mesons. $N_{lll}=N_{l}(N_{l}-1)(N_{l}-2)$ is the number of all possible
three quark combinations and $N_{l\bar{l}}=N_{l}N_{\bar{l}}$ is the
number of all possible light-flavor quark antiquark pairs. $N_{iter}$
is number of permutation for $l_{1}l_{2}l_{3}$ and is taken as 6
for three different flavors and 3 for two identical flavor and 1 for
three identical flavor, respectively. 

$C_{B_{j}}$ and $C_{M_{j}}$ are introduced to tune the production
weight of hadrons with same quark content but different spins. In
this paper, we only consider the ground state $J^{P}=0^{-},\,1^{-}$
mesons and $J^{P}=(1/2)^{+},\,(3/2)^{+}$ baryons in flavor SU(3)
group. We introduce a parameter $R_{V/P}$ to denote the relative
production weight of the vector mesons to the pseudoscalar mesons
with the same flavor composition. Then, we get $C_{M_{j}}=1/(1+R_{V/P})$
for $J^{P}=0^{-}$ mesons and $C_{M_{j}}=R_{V/P}/(1+R_{V/P})$ for
$J^{P}=1^{-}$ mesons. Similarly, we introduce a parameter $R_{D/O}$
to denote the relative production weight of the decuplet baryons to
the octet baryons with the same flavor composition. Then, we have
$C_{B_{j}}=1/(1+R_{D/O})$ for $J^{P}=(1/2)^{+}$ baryons and $C_{B_{j}}=R_{D/O}/(1+R_{D/O})$
for $J^{P}=(3/2)^{+}$ baryons, except $C_{\Lambda}=C_{\Sigma^{0}}=1/(2+R_{D/O}),~C_{\Sigma^{*0}}=R_{D/O}/(2+R_{D/O}),~C_{\Delta^{++}}=C_{\Delta^{-}}=C_{\Omega^{-}}=1$.
Here, $R_{V/P}$ and $R_{O/D}$ are set to be 0.45 and 0.5, respectively,
according to our recent work in $pp$ collisions at $\sqrt{s}=$13
 TeV \citep{Zhang:2018vyr}. 

Similar to Eqs. (\ref{prob_B_lf}) and (\ref{prob_M_lf}), the combination
probability of a charm quark and light-flavor (anti-)quark(s) can
be parameterized by 
\begin{eqnarray}
P_{cl_{1}l_{2}\rightarrow B_{j}} & = & C_{B_{j}}N_{iter}\frac{\overline{N}_{B_{c}}}{N_{cll}},\label{prob_B_hf}\\
P_{c\bar{l}_{1}\rightarrow M_{j}} & = & C_{M_{j}}\frac{\overline{N}_{M_{c}}}{N_{c\bar{l}}},\label{prob_M_hf}
\end{eqnarray}
where $N_{cll}=N_{c}N_{l}(N_{l}-1)$, $N_{c\bar{l}}=N_{c}N_{\bar{l}}$
and $N_{iter}$ equals to 1 as $l_{1}=l_{2}$ or 2 as $l_{1}\neq l_{2}$.
In this paper, we consider the ground state $J^{P}=0^{-},\,1^{-}$
single-charmed mesons, $J^{P}=\left(1/2\right)^{+}$ triplet and sextet
single-charmed baryons, and $J^{P}=\left(3/2\right)^{+}$ sextet single-charmed
baryons. Similar to light-flavor mesons, we introduce the parameter
$R'_{V/P}$ to denote the relative production weight of the vector
mesons to the pseudoscalar mesons. Different from light-flavor baryons,
we introduce two parameters in single-charmed baryons. We use a parameter
$R_{S1/T}$ to denote the relative production weight of $J^{P}=\left(1/2\right)^{+}$
sextet baryons to $J^{P}=\left(1/2\right)^{+}$ triplet baryons with
the same flavor composition, and another parameter $R_{S3/S1}$ to
denote that of $J^{P}=\left(3/2\right)^{+}$ sextet baryons to $J^{P}=\left(1/2\right)^{+}$
sextet baryons. We take $R'_{V/P}=1.5$, $R_{S1/T}=0.5$ and $R_{S3/S1}=1.4$
according to our previous work of single-charmed hadrons \citep{Li:2017zuj}.
We emphasize that yields and momentum spectra of final state charmed
baryons $\Lambda_{c}^{+}$, $\Xi_{c}^{0,+}$ and $\Omega_{c}^{0}$
after taking strong and electromagnetic decays into account are actually
insensitive to parameters $R_{S1/T}$ and $R_{S3/S1}$. 

The unitarity of the hadronization process constrains the number of
the formed hadrons, 
\begin{align}
\overline{N}_{M}+3\overline{N}_{B}+\overline{N}_{M_{\bar{c}}}+2\overline{N}_{B_{c}} & =N_{l},\label{eq:con_NL}\\
\overline{N}_{M}+3\overline{N}_{\bar{B}}+\overline{N}_{M_{c}}+2\overline{N}_{\bar{B}_{\bar{c}}} & =N_{\bar{l}},\label{eq:con_NLbar}\\
\overline{N}_{M_{c}}+\overline{N}_{B_{c}} & =N_{c},\label{eq:con_NC}\\
\overline{N}_{M_{\bar{c}}}+\overline{N}_{\bar{B}_{\bar{c}}} & =N_{\bar{c}},\label{eq:con_NCbar}
\end{align}
where we neglect the contribution of multi-charmed hadrons. Because
of small value for the relative production ratio $N_{c}/N_{l}\sim\mathcal{O}(1\%)$
in high energy $pp$, $p$A and AA collisions, we can neglect the
contribution of charmed hadrons in Eqs. (\ref{eq:con_NL}) and (\ref{eq:con_NLbar})
and then obtain the separate constraint for $N_{l}$ and $N_{c}$, respectively. 

In collisions at LHC energies, the approximation of charge conjugation symmetry
$N_{q_{i}}=N_{\bar{q}_{i}}$ and $N_{h}=N_{\bar{h}}$ is usually
satisfied. Therefore, the above unitarity constraints are reduced
to $\overline{N}_{M}+3\overline{N}_{B}\approx N_{l}$ and $\overline{N}_{M_{c}}+\overline{N}_{B_{c}}=N_{c}$.
 We can define the competition factor $R_{B/M}=\overline{N}_{B}/\overline{N}_{M}$
to quantify the production weight of baryons in light-flavor sector
and take it as a model parameter. Then we can calculate 
\begin{align}
\overline{N}_{B} & =\frac{R_{B/M}}{1+3R_{B/M}}N_{l},\label{eq:NB_tot}\\
\overline{N}_{M} & =\frac{1}{1+3R_{B/M}}N_{l}.\label{eq:NM_tot}
\end{align}
 We found that $R_{B/M}=0.087\pm0.04$ can well explain data of yield
densities of light-flavor hadrons in relativistic heavy-ion collisions
at RHIC and LHC energies and those in $pp$ and $p$Pb collisions
at LHC energies \citep{Song:2013isa,Shao:2017eok,Song:2020lpr}. We
also define a competition factor $R_{B/M}^{\left(c\right)}$ for single-charmed
hadrons and obtain
\begin{align}
\overline{N}_{B_{c}} & =\frac{R_{B/M}^{(c)}}{1+R_{B/M}^{(c)}}N_{c},\label{eq:NBc_tot}\\
\overline{N}_{M_{c}} & =\frac{1}{1+R_{B/M}^{(c)}}N_{c}.\label{eq:NMc_tot}
\end{align}
We found that $R_{B/M}^{\left(c\right)}$ is about $0.425\pm0.025$
in our recent works \citep{Song:2018tpv,Li:2017zuj} by fitting the
midrapidity data of $\Lambda_{c}^{+}$ in $pp$ collisions at $\sqrt{s}=7$
 TeV and those in $p$Pb collisions at $\sqrt{s_{NN}}=5.02$ TeV measured
by ALICE collaboration \citep{Acharya:2017kfy}. 

Quark momentum distributions $f_{q_{1}q_{2}q_{3}}(p_{1},p_{2},p_{3})$
and $f_{q_{1}\bar{q}_{2}}(p_{1},p_{2})$ are inputs of the model. When they are given, we can obtain $N_{q_1 q_2 q_3}$, $N_{q_1 \bar{q}_2}$, $N_{q_i}$ and $N_{\bar{q}_i}$ after integrating over momenta. Then, substituting Eqs.~(\ref{eq:NB_tot}-\ref{eq:NM_tot}) into Eqs.~(\ref{prob_B_lf}-\ref{prob_M_lf}) and subsequently substituting the latter into Eqs.~(\ref{eq:NBi_gen}-\ref{eq:NMi_gen}), we can obtain yields of light-flavor hadrons. By Eqs.~(\ref{eq:NBi_gen}-\ref{eq:NMi_gen}) and Eqs.~(\ref{eq:coe_ABi}-\ref{eq:coe_AMi}), we can calculate coefficients $\kappa_{B_j}$ and $\kappa_{M_j}$. Substituting them into Eqs.~(\ref{eq:fbi}-\ref{eq:fmi}), we can obtain momentum distributions of light-flavor hadrons.  
Calculations of single-charmed hadron are similar. 

We also consider the physical situation that the numbers of quarks
are not fixed values but are fluctuated event by event in high energy
collisions. As we did in Ref. \citep{Shao:2017eok}, we consider the
Poisson distribution as the base line to simulate the numbers of quarks
of different kinds of flavors produced in midrapidity range in each
event. Then we take the event average of the numbers of hadrons to
obtain their yield densities. We note that the effect of quark number
fluctuations influences little on production of mesons and weakly
on that of baryons containing up and down quarks but obviously on
that of multi-strange baryons such as $\Omega$ \citep{Shao:2017eok}.
The fluctuations for momentum distributions of quarks are not considered
at the moment. 

We finally consider the decay effects of short-life hadrons on production
of stable hadrons, 
\begin{equation}
f_{h_{j}}^{(final)}(p)=f_{h_{j}}(p)+\sum_{i\neq j}\int dp'f_{h_{i}}(p')D_{ij}(p',p),
\end{equation}
where the decay function $D_{ij}(p',p)$ is calculated by decay kinetics
and decay branch ratios reported by Particle Data Group \citep{Agashe:2014kda}. 

As a short summary of this section, we emphasize that the EVC model
is essentially a statistical model based on the constituent quark
degrees of freedom at hadronization. In deriving momentum spectra
and yields of hadrons, stochastic feature of quark combination and
flavor-independence of strong interactions are mainly used. The effect
of flavor symmetry broken is taken into account, on the one hand,
by the difference in momentum distributions (and also numbers) of
quarks with different flavors which will be discussed in the following
text, and on the other hand, by the flavor-dependent parameter such
as the difference between $R_{B/M}$ in light-flavors and $R_{B/M}^{(c)}$
in charms. In addition, non-perturbative dynamics in combination process
which are difficult to be calculated in first principles are parameterized
in the model. We expect that values of these parameters such as $R_{V/P}$
and $R_{B/M}$ are stable in different high energy collisions, as
indicated by our available studies up to now. Finally, the momentum
distribution $dN_{h,q}/dp$ is a general denotation. In this paper, we focus on the transverse production
of hadrons at midrapidity, then the momentum distribution $dN_{h,q}/dp$
refers to $dN_{h,q}/dp_{T}dy$ at midrapidity. 

\section{Results of light-flavor hadrons\label{sec:LightFlavor_Results}}
In our model, momentum distributions of light-flavor constituent quarks at hadronization are inputs. Because they are difficult to be calculated in the low $p_T$ range from first principles, we determine them by fitting experimental data of identified hadrons in our model. Considering that the available experimental measurements are mainly inclusive distribution, here we assume the factorization approximation for the joint momentum distribution of (anti-)quarks, i.e.,~$f_{q_{1}q_{2}q_{3}}\left(p_{T_{1}},p_{T_{2}},p_{T_{3}}\right)=f_{q_{1}}\left(p_{T_{1}}\right)f_{q_{2}}\left(p_{T_{2}}\right)f_{q_{3}}\left(p_{T_{3}}\right)$ and $f_{q_{1}\bar{q}_{2}}\left(p_{T_{1}},p_{T_{2}}\right)=f_{q_{1}}\left(p_{T_{1}}\right)f_{\bar{q}_{2}}\left(p_{T_{2}}\right)$.  In addition, we take the isospin symmetry $f_{u}\left(p_{T}\right)=f_{d}\left(p_{T}\right)$ and the charge-conjugation symmetry $f_{q_{i}}\left(p_{T}\right)=f_{\bar{q}_{i}}\left(p_{T}\right)$ for $p_{T}$ spectra of (anti-)quarks at midrapidity at LHC energies. Finally, we have only two inputs $f_u(p_T)$ and $f_s(p_T)$ in light-flavor sector which can be conveniently determined by experimental data of a few of hadrons. 

In this section, we study the production of light-flavor hadrons in the low and intermediate $p_T$ range at
midrapidity in $pp$ collisions at $\sqrt{s}=5.02$ TeV. In particular,
we discuss a quark number scaling property for $p_{T}$ spectra of
$\Omega^-$ and $\phi$ and study the $p_{T}$ dependence of $\Omega/\phi$
ratio. We also study the property of the extracted $p_{T}$ spectra
of up quarks and strange quarks. 
\subsection{Scaling property for $p_{T}$ spectra of $\Omega^-$ and $\phi$}

In this subsection, we discuss an interesting correlation between
$p_{T}$ spectrum of $\Omega^-$ and that of $\phi$, which gives a
first insight into hadron production mechanism at hadronization. $\Omega$
and $\phi$ consist of strange quarks/antiquarks, exclusively. In
EVC model, $p_{T}$ spectra of $\Omega^-$ and $\phi$ have simple expressions
\begin{align}
f_{\Omega}\left(3p_{T}\right) & =\kappa_{\Omega}\left[f_{s}\left(p_{T}\right)\right]^{3},\label{eq:fpt_Omega}\\
f_{\phi}\left(2p_{T}\right) & =\kappa_{\phi}f_{s}\left(p_{T}\right)f_{\bar{s}}\left(p_{T}\right)=\kappa_{\phi}\left[f_{s}\left(p_{T}\right)\right]^{2},\label{eq:fpt_phi}
\end{align}
where we use $f_{s}\left(p_{T}\right)=f_{\bar{s}}\left(p_{T}\right)$
for midrapidity at LHC energy. We then obtain the following correlation 

\begin{equation}
f_{\phi}^{1/2}\left(2p_{T}\right)=\kappa_{\phi,\Omega}f_{\Omega}^{1/3}\left(3p_{T}\right)\label{eq:qns}
\end{equation}
where the coefficient $\kappa_{\phi,\Omega}=\kappa_{\phi}^{1/2}/\kappa_{\Omega}^{1/3}$
is independent of $p_{T}$. Eq. (\ref{eq:qns}) means that, in the
stochastic combination scenario of quarks and antiquarks at hadronization,
$p_{T}$ spectra of $\Omega^-$ and $\phi$ have a strong correlation
based on the number of strange (anti-)quarks they contain. Therefore,
we call Eq. (\ref{eq:qns}) the quark number scaling property. 

\begin{figure}[h]
\includegraphics[scale=0.35]{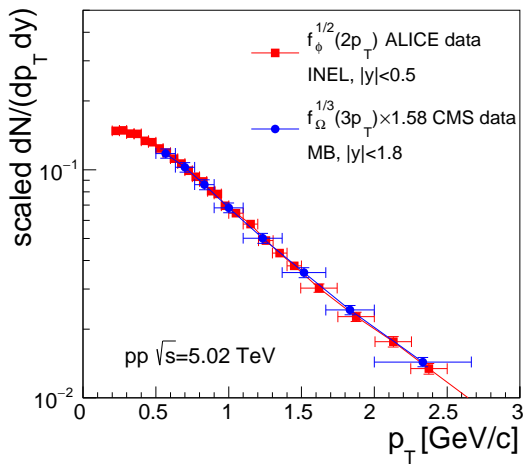}\caption{The scaled $p_{T}$ spectra of $\Omega$ and $\phi$ in $pp$ collisions
at $\sqrt{s}=5.02$ TeV. Experimental data of $\Omega$ and $\phi$
are from \citep{Tripathy:2018ehz,Sirunyan:2019rfz}.\label{fig:qns}}
\end{figure}

In Fig. \ref{fig:qns}, we test Eq. (\ref{eq:qns}) by the preliminary
data of $p_{T}$ spectrum of $\phi$ in the rapidity interval $|y|<0.5$
in inelastic events in $pp$ collisions at $\sqrt{s}=5.02$ TeV measured
by ALICE collaboration \citep{Tripathy:2018ehz} and data of $\Omega$
(i.e., $\Omega^{-}+\bar{\Omega}^{+}$) in the rapidity interval $|y|<1.8$
in minimum-bias events measured by CMS collaboration \citep{Sirunyan:2019rfz}.
In order to compare the scaled data from two different collaborations, 
the coefficient $\kappa_{\phi,\Omega}$ is taken as 1.58 but not the direct calculation of our model\footnotemark[1]. We see that
the scaled data of $\Omega$ are in good agreement with those of $\phi$.
Furthermore, we know from Eqs. (\ref{eq:fpt_Omega}) and (\ref{eq:fpt_phi})
that Eq. (\ref{eq:qns}) equals to $f_{s}\left(p_{T}\right)$ multiplying
by a $p_{T}$-independent coefficient $\sqrt{\kappa_{\phi}}$. Therefore,
Fig. \ref{fig:qns} also gives the direct information on the $p_{T}$
spectrum of strange quarks at hadronization in $pp$ collisions at
$\sqrt{s}=5.02$ TeV.

\subsection{$p_{T}$ spectra of $p$, $K^{*0}$, $\Lambda$ and $\Xi$}

    We parameterize the $p_{T}$ spectrum of quarks by a L\'evy-Tsallis functional form \cite{Tsallis:1987eu}. According to Eq.~(\ref{eq:fpt_phi}),  we can use our model to fit the scaled data of $\phi$ in Fig. \ref{fig:qns} to  obtain the $p_{T}$ spectrum of strange quarks $f_{s}\left(p_{T}\right)$ at hadronization. 
We further use the model to fit experimental data of $p_{T}$ spectrum of proton \citep{Acharya:2019yoi} to obtain the $p_{T}$ spectrum of up/down quarks $f_{u}\left(p_{T}\right)$ at hadronization.  The properties of $f_s(p_T)$ and $f_u(p_T)$ will be discussed in Sec.~\ref{subsec:fus_diff}.

\begin{figure}[h]
\includegraphics[scale=0.43]{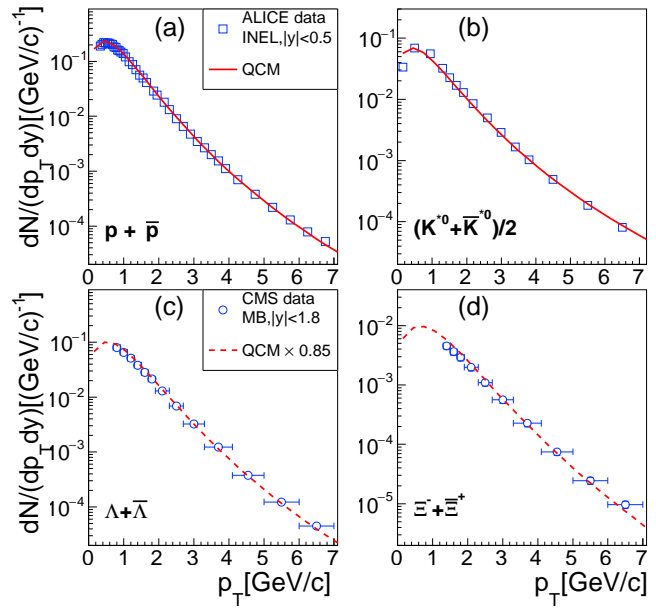}\caption{$p_{T}$ spectra of $p+\bar{p}$, $\left(K^{*0}+\bar{K}^{*0}\right)/2$,
$\Lambda+\bar{\Lambda}$ and $\Xi^{-}+\bar{\Xi}^{+}$ in $pp$ collisions
at $\sqrt{s}=5.02$ TeV. Symbols are experimental data \citep{Garg:2018fwt,Acharya:2019yoi,Sirunyan:2019rfz} and lines with label ``QCM" are model results. \label{fig:fpt_p_Kstar_Lam_Xi}}
\end{figure}

When $f_u(p_T)$ and $f_s(p_T)$ are obtained, we can calculate $p_{T}$ spectra of various light-flavor hadrons. 
In Fig. \ref{fig:fpt_p_Kstar_Lam_Xi}(b), we show the result for $p_T$ spectrum of $\left(K^{*0}+\bar{K}^{*0}\right)/2$ and compare it with the experimental data \citep{Garg:2018fwt}.  We see a good agreement. Note that data of
$\phi$, proton and $K^{*}$ are all ALICE data in inelastic events
and rapidity interval $|y|<0.5$. In Fig. \ref{fig:fpt_p_Kstar_Lam_Xi} (c) and (d), we present
results of $\Lambda+\bar{\Lambda}$ and $\Xi^{-}+\bar{\Xi}^{+}$ and
compare them with experimental data of CMS collaboration \citep{Sirunyan:2019rfz}.
Because CMS experiments select the minimum-bias events and rapidity
interval $|y|<1.8$ which are different from ALICE experiments, we
multiply our results of $\Lambda$ and $\Xi$ by a constant 0.85 to
test the shape of $p_{T}$ distributions of
hyperons predicted in our model \footnotemark[1]. We see a good description for the shape of $p_{T}$ spectra
of two hyperons. 

\footnotetext[1]{By examining the available experimental data for $p_{T}$ spectra
of hyperons in $pp$ collisions at $\sqrt{s}=7$ TeV measured by ALICE
collaboration and those by CMS collaboration \citep{Khachatryan:2011tm,Abelev:2014qqa,Abelev:2012jp},
we notice that the average transverse momentum $\langle p_{T}\rangle$
and the shape of $p_{T}$ distributions measured by two collaborations
are quite consistent, although the center values of $dN/dy$ measured
by two collaborations have a certain difference. Therefore, in this
paper, we put two data sets in $pp$ collisions at $\sqrt{s}=$ 5.02 
 TeV into together to test our model.}

\subsection{Ratio $\Omega/\phi$ as the function of $p_{T}$}

\begin{figure}[h]
\includegraphics[scale=0.45]{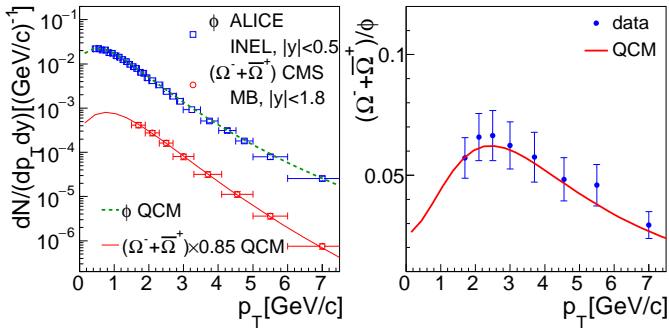}\caption{(a) $p_{T}$ spectra of $\Omega$ and $\phi$ in $pp$ collisions
at $\sqrt{s}=5.02$ TeV. (b) ratio $\Omega/\phi$ as the function
of $p_{T}$. Lines with label ``QCM" are model results and symbols are experimental
data \citep{Tripathy:2018ehz,Sirunyan:2019rfz}.\label{fig:ratio_Omg_phi}}
\end{figure}

In Fig. \ref{fig:ratio_Omg_phi}(a), we show the fitting result for
$p_{T}$ spectrum of $\phi$ and the calculation result for $p_{T}$
spectrum of $\Omega$ (i.e., $\Omega^{-}+\bar{\Omega}^{+}$) in $pp$
collisions at $\sqrt{s}=5.02$ TeV. Similar to the case of $\Lambda$
and $\Xi$ in Fig. \ref{fig:fpt_p_Kstar_Lam_Xi}, we also multiply
our result of $\Omega$ by a constant 0.85 in order to compare with
the shape of experimental data for $p_{T}$ spectrum of $\Omega$
measured by CMS collaboration \citep{Sirunyan:2019rfz}. As indicated
by the quark number scaling property in Fig. \ref{fig:qns}, we see
that $p_{T}$ spectra of $\phi$ and $\Omega$ can be simultaneously
described by our model. 

In Fig. \ref{fig:ratio_Omg_phi}(b), we show the result for $\Omega/\phi$
ratio as the function of $p_{T}$ and compare with experimental data.
Here, the experimental data for $\Omega/\phi$ ratio are calculated
by data of their inclusive $p_{T}$ spectra in Fig. \ref{fig:ratio_Omg_phi}
(a) with the propagation of statistical uncertainties \citep{Tripathy:2018ehz,Sirunyan:2019rfz}.
We see that ratio $\Omega/\phi$ firstly increases with $p_{T}$ in
the low $p_{T}$ range ($p_{T}\lesssim2.5$ GeV/c) and then turns
to decrease with $p_{T}$ at larger $p_{T}$. Our model result, the
solid line, well explains experimental data. 

This non-monotonic $p_{T}$ dependence of the ratio of baryon to meson
and, in particular, the enhancement in the low $p_{T}$ range have
been observed many times in relativistic heavy-ion collisions \citep{Abelev:2008aa,Aggarwal:2010ig,Abelev:2013xaa,Adamczyk:2015lvo,Adam:2019koz}
and in $pp$ and $p$Pb collisions at LHC energies \citep{Khachatryan:2011tm,Abelev:2012hy,Abelev:2013haa}.
This typical behavior of baryon to meson ratio is usually regarded as
the consequence of the quark combination mechanism at hadronization
\citep{Greco:2003xt,Fries:2003vb,Hwa:2002tu,Chen:2006vc,Shao:2009uk}.
In this paper, taking $\Omega/\phi$ ratio as an example, we carry
out a simple derivation to further clarify the underlying physics
of such $p_{T}$ dependence of baryon to meson ratios in the low and
intermediate $p_{T}$ range. 

To understand the $p_{T}$ dependence of $\Omega/\phi$ ratio, we
calculate the slop of the ratio 

\begin{align}
 & \left[\frac{f_{\Omega}\left(p_{T}\right)}{f_{\phi}\left(p_{T}\right)}\right]^{'}\nonumber \\
 & =\frac{f_{\Omega}\left(p_{T}\right)}{f_{\phi}\left(p_{T}\right)}\left[\frac{f_{\Omega}^{'}\left(p_{T}\right)}{f_{\Omega}\left(p_{T}\right)}-\frac{f_{\phi}^{'}\left(p_{T}\right)}{f_{\phi}\left(p_{T}\right)}\right]\nonumber \\
 & =\frac{f_{\Omega}\left(p_{T}\right)}{f_{\phi}\left(p_{T}\right)}\left[\frac{\partial\ln\left(f_{s}\left(p_{T}/3\right)\right)}{\partial\left(p_{T}/3\right)}-\frac{\partial\ln\left(f_{s}\left(p_{T}/2\right)\right)}{\partial\left(p_{T}/2\right)}\right].
\end{align}
Using the mean-value theorem, the term in bracket in the last line
becomes 
\begin{align}
 & \frac{\partial\ln\left(f_{s}\left(p_{T}/3\right)\right)}{\partial\left(p_{T}/3\right)}-\frac{\partial\ln\left(f_{s}\left(p_{T}/2\right)\right)}{\partial\left(p_{T}/2\right)}\nonumber \\
 & =-\frac{1}{6}p_{T}\left[\ln f_{s}\left(\xi\right)\right]^{''}
\end{align}
with $p_{T}/3<\xi<p_{T}/2$. Finally, we have 
\begin{equation}
\left[\ln\frac{f_{\Omega}\left(p_{T}\right)}{f_{\phi}\left(p_{T}\right)}\right]^{'}=-\frac{1}{6}p_{T}\left[\ln f_{s}\left(\xi\right)\right]^{''},
\end{equation}
which means that the slop of the $\Omega/\phi$ ratio is influenced
by the second derivative of the logarithm of strange quark distribution. 

The second derivative of a distribution is related to that this distribution
is convex or concave in shape. This can be conveniently read from
Fig.~\ref{fig:qns} or Fig.~\ref{fig:fu_fs}. We see that $\left[\ln f_{s}(p_{T,s})\right]^{''}<0$
as $p_{T,s}\lesssim0.9$ GeV/c and $\left[\ln f_{s}(p_{T,s})\right]^{''}>0$
as $1.0\lesssim p_{T,s}\lesssim2.5$ GeV/c. Therefore, the $\Omega/\phi$
ratio increases with $p_{T}$ in the range $p_{T}\lesssim2-3$ GeV/c
and decreases with $p_{T}$ at larger $p_{T}$.

As we know, quarks of small $p_T$ mainly come from soft QCD process and $p_T$ distribution of these quarks is usually described by a thermal-like function $\exp[-\sqrt{p_{T}^{2}+m^{2}}/T]$ which just has the property $\left[\ln f_{s}\left(p_{T,s})\right)\right]^{''}<0$ leading to the increase of $\Omega/\phi$ ratio.
Quarks of large $p_T$ mainly come from hard QCD process and $p_T$ distribution of these quarks is usually described by a jet-like function  $(1+p_{T}/p_{0})^{-n}$ with $p_{0}>0$ and $n>0$ which just has the property $\left[\ln f_{s}\left(p_{T,s}\right)\right]^{''}>0$ leading to the decrease of $\Omega/\phi$ ratio.
Therefore, we emphasize that the observed non-monotonic $p_{T}$
dependence of the $\Omega/\phi$ ratio not only depends on quark combination
mechanism but also depends on the property of the momentum distribution
of strange quarks at hadronization. 

\subsection{Difference between $u$ and $s$ quarks in $p_{T}$ spectrum \label{subsec:fus_diff}}

\begin{figure}[h]
\includegraphics[scale=0.45]{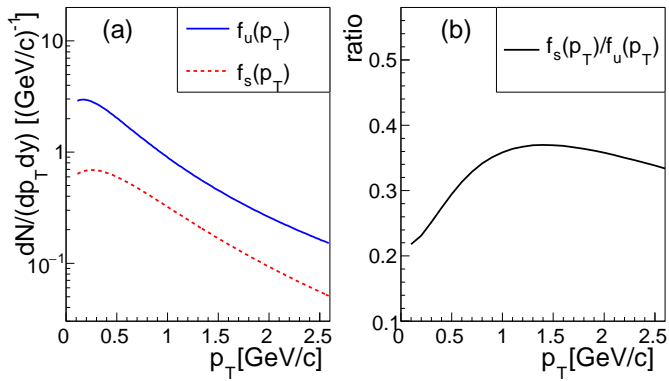}\caption{(a) $p_{T}$ spectra of up and strange quarks at hadronization in
inelastic $pp$ collisions at $\sqrt{s}=5.02$ TeV. (b) the ratio
$f_{s}\left(p_{T}\right)/f_{u}\left(p_{T}\right)$.\label{fig:fu_fs}}
\end{figure}

In Fig. \ref{fig:fu_fs}(a), we show the $p_{T}$ spectra of up and
strange quarks at hadronization extracted from data of $\phi$ and
proton in inelastic $pp$ collisions at $\sqrt{s}=5.02$ TeV. The
ratio in $p_{T}$-integrated yield density between strange quarks
and up quarks, i.e., strangeness suppression factor, 
\begin{equation}
\lambda_{s}=\frac{dN_{s}/dy}{dN_{u}/dy}
\end{equation}
 is about 0.3. In panel (b), we show the spectrum ratio of strange
quarks to up quarks. We see that the ratio increases with $p_{T}$
as $p_{T}\lesssim1$ GeV/c and turns to weakly decrease at larger
$p_{T}$. We note that this property is also observed in $pp$ collisions
at other collision energies and in relativistic heavy-ion collisions
\citep{Song:2017gcz,Zhang:2018vyr,Song:2020lpr}. 

\begin{figure}[h]
\includegraphics[scale=0.35]{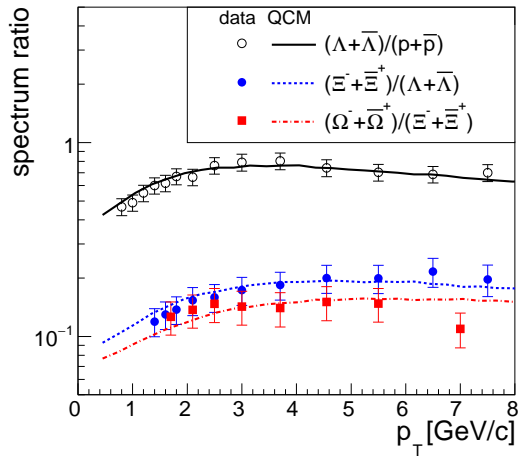}\caption{Spectrum ratios $\left(\Lambda+\bar{\Lambda}\right)/\left(p+\bar{p}\right)$,
$(\Xi^{-}+\bar{\Xi}^{+})/\left(\Lambda+\bar{\Lambda}\right)$, and
$\left(\Omega^{-}+\bar{\Omega}^{+}\right)/\left(\Xi^{-}+\bar{\Xi}^{+}\right)$
as the function of $p_{T}$. Experimental data are from \citep{Acharya:2019yoi,Sirunyan:2019rfz}.
\label{fig:ls_cascade}}
\end{figure}

The difference in the $p_{T}$ spectrum between up quarks and strange
quarks will influence $p_{T}$ spectra of the formed hadrons with
different strange quark content. In Fig. \ref{fig:ls_cascade}, we
show ratios $\left(\Lambda+\bar{\Lambda}\right)/\left(p+\bar{p}\right)$,
$\left(\Xi^{-}+\bar{\Xi}^{+}\right)/\left(\Lambda+\bar{\Lambda}\right)$,
and $\left(\Omega^{-}+\bar{\Omega}^{+}\right)/\left(\Xi^{-}+\bar{\Xi}^{+}\right)$
as the function of $p_{T}$. Symbols are experimental data and different
kinds of lines are model results. The experimental data of three ratios
are calculated by data of their inclusive $p_{T}$ spectra 
with propagation of statistical uncertainties \citep{Acharya:2019yoi,Sirunyan:2019rfz}.
We see that data of three ratios in the low $p_{T}$ range ($p_{T}\lesssim4$
GeV/c) all increase with $p_{T}$. In our model, this is because of
the quark level property shown in Fig. \ref{fig:fu_fs}(b) as $p_{T,q}\lesssim1.3$
GeV/c. The hierarchy in magnitude for data of three ratios can be
understood in our model by ratios of yield densities, 
\begin{align}
\frac{dN_{\Lambda}/dy}{dN_{p}/dy} & \approx\frac{7.7}{4}\lambda_{s},\\
\frac{dN_{\Xi}/dy}{dN_{\Lambda}/dy} & \approx\frac{3}{7.7}\lambda_{s},\\
\frac{dN_{\Omega}/dy}{dN_{\Xi}/dy} & \approx\frac{1}{3}\lambda_{s},
\end{align}
where coefficients before $\lambda_{s}$ are due to the iteration
factor $N_{iter}$ in Eq. (\ref{prob_B_lf}) and strong/electromagnetic
decay contribution of decuplet baryons, see \citep{Wang:2012cw,Zhang:2018vyr}
for the detailed analytical expressions of their yields. 

\section{Results of charmed hadrons \label{sec:HF_results}}

In this section, we study the production of single-charmed hadrons
in $pp$ collisions at $\sqrt{s}=5.02$ TeV. We firstly extract the
$p_{T}$ spectrum of charm quarks and compare it with the calculation
result of perturbative QCD method. Then, we present results of $D$
mesons and $\Lambda_{c}^{+}$ baryon, and we compare them with experimental
data. We also predict the $p_{T}$-differential cross-section of $\Xi_{c}^{0,+}$
and $\Omega_{c}^{0}$, and their ratios to $D$ mesons as the function
of $p_{T}$. 

\subsection{$p_{T}$ spectrum of charm quarks \label{subsec:fc}}

\begin{figure}[h]
\includegraphics[width=0.98\columnwidth]{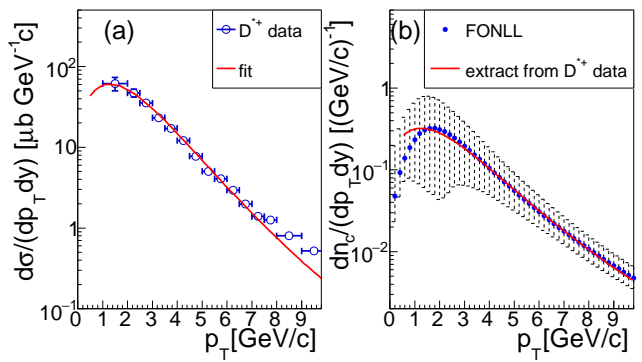}\caption{(a) Fit to data of $D^{*+}$ in EVC model. (b) Comparison between
the normalized $p_{T}$ distribution of charm quarks obtained in our
model and calculation in FONLL \citep{Cacciari:1998it,Cacciari:2001td}.
\label{fig:fc}}
\end{figure}

In Fig. \ref{fig:fc}(a), we apply the EVC model to fit experimental
data for differential cross-section of $D^{*+}$ in $pp$ collisions
at $\sqrt{s}=5.02$ TeV. In the fit, we have used the $p_{T}$ spectrum
of $u$ quarks obtained in previous section and then obtain the $p_{T}$
distribution of charm quarks at hadronization. In panel (b), we normalize
the obtained charm quark distribution and compare it with the perturbative QCD calculation
in Fixed-Order Next-to-Leading-Logarithmic (FONLL) scheme \citep{Cacciari:1998it,Cacciari:2001td}.
We find a good consistency in the studied $p_{T}$ range within theoretical
uncertainties. 
By fitting $D^{*+}$ data, we obtain the $p_{T}$ integrated
cross-section of charm quarks $d\sigma_{c}/dy=1.0$ mb. This value
is higher than the center value of default FONLL calculation $0.461_{-0.31}^{+0.58}$
mb \citep{Cacciari:1998it,Cacciari:2001td} but is still located in
its theoretical uncertainties. 

\subsection{Results of single-charmed hadrons }

Using the extracted $p_{T}$ spectrum of charm quarks in the above
subsection and those of light-flavor quarks in Sec. \ref{subsec:fus_diff},
we can calculate $p_{T}$ spectra of other single-charmed hadrons.
In Table \ref{tab:charm_hadron_yield}, we firstly present the $p_{T}$-integrated
cross-section $d\sigma/dy$ of $D^{0,+}$, $D_{s}^{+}$, $\Lambda_{c}^{+}$,
$\Xi_{c}^{0,+}$ and $\Omega_{c}^{0}$ in inelastic $pp$ collisions
at $\sqrt{s}=5.02$ TeV. We firstly list analytical expressions and then numerical
results as $d\sigma_{c}/dy=1.0\,mb$ and compare with latter with available experimental data \cite{Acharya:2019mgn,Acharya:2020lrg}. 

\begin{table}[H]
\centering{}\caption{$d\sigma/dy$ of single-charmed hadrons in EVC model as $d\sigma_{c}/dy=1.0\,mb$. Strong and electromagnetic decay contributions from other
single-charmed hadrons in ground-state have been included. Experimental data are from \cite{Acharya:2019mgn,Acharya:2020lrg}.\label{tab:charm_hadron_yield}}
\begin{tabular*}{\linewidth}{@{\extracolsep{\fill}}cccc}
\toprule 
    $\frac{d\sigma}{dy}$ & analytical & numerical ($\mu b$) & data ($\mu b$)\tabularnewline
\midrule
\midrule 
    $D^{0}$ & $\frac{1+1.677R'_{V/P}}{1+R'_{V/P}}\frac{1}{2+\lambda_{s}}\frac{1}{1+R_{B/M}^{(c)}}\frac{d\sigma_{c}}{dy}$ & 429  &  $447\pm20$\tabularnewline
\midrule 
    $D^{+}$ & $\frac{1+0.323R'_{V/P}}{1+R'_{V/P}}\frac{1}{2+\lambda_{s}}\frac{1}{1+R_{B/M}^{(c)}}\frac{d\sigma_{c}}{dy}$ & 181     & $184\pm13$\tabularnewline
\midrule 
    $D_{s}^{+}$ & $\frac{\lambda_{s}}{2+\lambda_{s}}\frac{1}{1+R_{B/M}^{(c)}}\frac{d\sigma_{c}}{dy}$ & 91.5  & $95\pm9$\tabularnewline
\midrule 
    $\Lambda_{c}^{+}$ & $\frac{4}{\left(2+\lambda_{s}\right)^{2}}\frac{R_{B/M}^{(c)}}{1+R_{B/M}^{(c)}}\frac{d\sigma_{c}}{dy}$ & 225 & $230\pm16$\tabularnewline
\midrule 
    $\Xi_{c}^{0}$ & $2\frac{\lambda_{s}}{\left(2+\lambda_{s}\right)^{2}}\frac{R_{B/M}^{(c)}}{1+R_{B/M}^{(c)}}\frac{d\sigma_{c}}{dy}$ & 33.8 & \tabularnewline
\midrule 
    $\Xi_{c}^{+}$ & $2\frac{\lambda_{s}}{\left(2+\lambda_{s}\right)^{2}}\frac{R_{B/M}^{(c)}}{1+R_{B/M}^{(c)}}\frac{d\sigma_{c}}{dy}$ & 33.8 &\tabularnewline
\midrule 
    $\Omega_{c}^{0}$ & $\frac{\lambda_{s}^{2}}{\left(2+\lambda_{s}\right)^{2}}\frac{R_{B/M}^{(c)}}{1+R_{B/M}^{(c)}}\frac{d\sigma_{c}}{dy}$ & 5.07 &\tabularnewline
\bottomrule
\end{tabular*}
\end{table}

\begin{figure}[h]
\centering{}\includegraphics[scale=0.42]{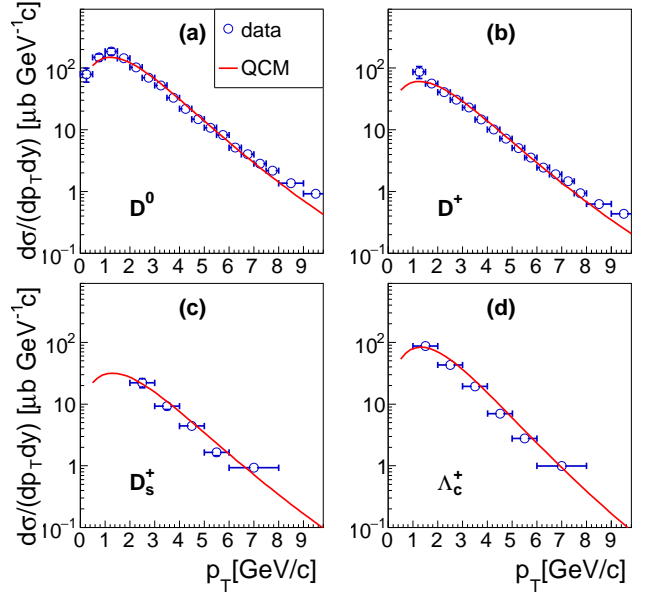}\caption{$p_{T}$ spectra of $D^{0,+}$, $D_{s}^{+}$ and $\Lambda_{c}^{+}$
at midrapidity in inelastic $pp$ collisions at $\sqrt{s}=5.02$ TeV.
Lines are results of EVC model and symbols are experimental data \citep{Acharya:2020lrg,Acharya:2019mgn}.
\label{fig:fhc}}
\end{figure}

In Fig. \ref{fig:fhc}, we present results for $p_{T}$ spectra of
$D^{0,+}$, $D_{s}^{+}$ and $\Lambda_{c}^{+}$ in inelastic $pp$
collisions at $\sqrt{s}=5.02$ TeV and compare them with experimental
data of ALICE collaboration \citep{Acharya:2020lrg,Acharya:2019mgn}.
We find a good agreement for these four hadrons in the low $p_{T}$
range ($p_{T}\lesssim7$ GeV/c). At larger transverse momentum $p_{T}\gtrsim8$
GeV/c, results for $D^{0,+}$ in our model are lower than experimental
data to a certain extent. 
This under-estimation maybe indicate the increased importance of fragmentation mechanism for charm quark hadronization at large $p_T$.

\begin{figure}[h]
\centering{}\includegraphics[scale=0.42]{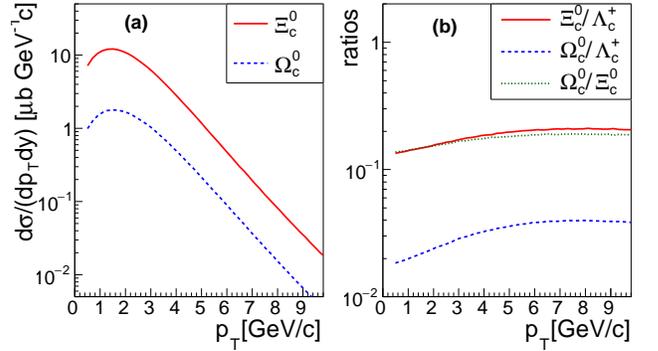}\caption{(a) $p_{T}$ spectra of $\Xi_{c}^{0}$ and $\Omega_{c}^{0}$ at midrapidity
in inelastic $pp$ collisions at $\sqrt{s}=5.02$ TeV. (b) Ratios
among charmed baryons as the function of $p_{T}$. \label{fig:fpt_Xic_Omg_c}}
\end{figure}

In Fig. \ref{fig:fpt_Xic_Omg_c} (a), we predict $p_{T}$ spectra
of $\Xi_{c}^{0}$ and $\Omega_{c}^{0}$ in inelastic $pp$ collisions
at $\sqrt{s}=5.02$ TeV.   
In comparison with the production of $\Lambda_{c}^{+}$, the production of $\Xi_{c}^{0}$
and that of $\Omega_{c}^{0}$ are suppressed due to the cascade strangeness. 
As shown by their $p_{T}$-integrated cross-sections in Table \ref{tab:charm_hadron_yield}, we have 
\begin{equation}
    \Lambda_{c}^{+}:\Xi_{c}^{0}:\Omega_{c}^{0}=1:\frac{1}{2}\lambda_{s}:\frac{1}{4}\lambda_{s}^{2}, \label{eq:LamXiOmg_c}
\end{equation}
Because $\lambda_s\approx0.3$ in inelastic $pp$ collisions, we see in Fig.~\ref{fig:fpt_Xic_Omg_c} (b) that ratio $\Xi_{c}^{0}/\Lambda_{c}^{+}$ is about 0.1-0.2 and $\Omega_{c}^{0}/\Lambda_{c}^{+}$ is about 0.02-0.03 in the low and intermediate $p_T$ range. Ratio $\Omega_{c}^{0}/\Xi_{c}^{0}$ is also the order of $\lambda_{s}/2$ and therefore is close to $\Xi_{c}^{0}/\Lambda_{c}^{+}$. In addition, we see that three ratios all increase with $p_T$ in the low $p_T$ range, which is because the difference between $f_s(p_T)$ and $f_u(p_T)$ as shown in Fig.~\ref{fig:fu_fs}.

\begin{figure}[h]
\centering{}\includegraphics[scale=0.44]{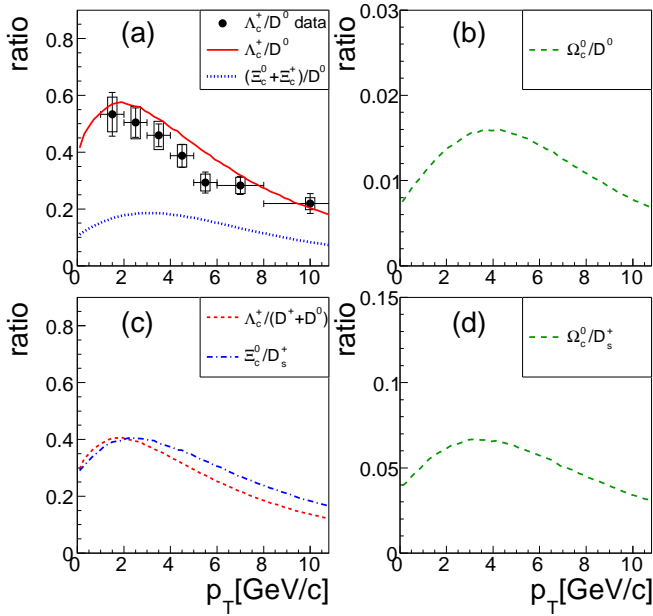}\caption{Ratios of charmed baryons to charmed mesons as the function of $p_{T}$
at midrapidity in inelastic $pp$ collisions at $\sqrt{s}=5.02$ TeV.\label{fig:BM_ratio_charm}}
\end{figure}

The ratio of baryon to meson as the function of $p_{T}$ is sensitive
to the production mechanism of hadrons at hadronization. In Fig. \ref{fig:BM_ratio_charm},
we show results for ratios of charmed baryons to charmed mesons as
the function of $p_{T}$ in $pp$ collisions at $\sqrt{s}=5.02$ TeV.
In Fig. \ref{fig:BM_ratio_charm} (a), we firstly show result of $\Lambda_{c}^{+}/D^{0}$
as the solid line. Comparing with experimental data of $\Lambda_{c}^{+}/D^{0}$
\citep{Acharya:2020lrg}, we see that our model result can well explain
the decreasing behavior of experimental data at $p_{T}\gtrsim2$ GeV/c.
In the low $p_{T}$ range ($p_{T}\lesssim2$ GeV/c), the ratio in
our model increases with the increasing $p_{T}$. This behavior can
be tested in the future as experimental data in this $p_{T}$ range
are richer and more precise. We note that experimental data of $p$Pb
and Pb-Pb collisions at small $p_{T}$ have indicated this property \citep{Acharya:2020lrg}. 

We further present result for $\left(\Xi_{c}^{0}+\Xi_{c}^{+}\right)/D^{0}$
in Fig. \ref{fig:BM_ratio_charm}(a) and that for $\Omega_{c}^{0}/D^{0}$
in Fig. \ref{fig:BM_ratio_charm}(b). We see that the magnitude of
$\left(\Xi_{c}^{0}+\Xi_{c}^{+}\right)/D^{0}$ at $p_{T}\approx 3$ GeV/c
is about 0.16 and that of $\Omega_{c}^{0}/D^{0}$ is only about 0.015.
This hierarchy property is due to the cascade strangeness suppression as shown in Eq.~(\ref{eq:LamXiOmg_c}). 

In order to reduce the effect of strangeness suppression and the dependence
of model parameters in these baryon to meson ratios, we propose two
new ratios $\Lambda_{c}^{+}/\left(D^{0}+D^{+}\right)$ and $\Xi_{c}^{0}/D_{s}^{+}$.
From Table \ref{tab:charm_hadron_yield}, two ratios in $p_{T}$-integrated
cross-sections are of the same magnitude
\begin{align}
\frac{d\sigma_{\Lambda_{c}^{+}}/dy}{d\sigma_{D^{0}+D^{+}}/dy} & =\frac{d\sigma_{\Xi_{c}^{0}}/dy}{d\sigma_{D_{s}^{+}}/dy}=\frac{2}{2+\lambda_{s}}R_{B/M}^{(c)}.
\end{align}
They are independent of model parameter $R'_{V/P}$, which is different
from $\Lambda_{c}^{+}/D^{0}$ ratio. They are also insensitive to
$\lambda_{s}$ since the change of $\lambda_{s}$ (e.g., 0.3-0.33)
only causes little (1\%) influence on two ratios. Finally, two ratios
directly relate to the production competition of baryon to meson in
charm sector which is characterized by the parameter $R_{B/M}^{(c)}$
in our model. Therefore, we propose these two ratios as direct observables
of baryon production weight in charm sector. Two ratios as the function
of $p_{T}$ are shown in Fig. \ref{fig:BM_ratio_charm}(c) where we
see they are close to each other. 

In Fig. \ref{fig:BM_ratio_charm}(d), we show result for $\Omega_{c}^{0}/D_{s}^{+}$
as the function of $p_{T}$. As indicated by their ratio in $p_{T}$-integrated
cross-section
\begin{equation}
\frac{d\sigma_{\Omega_{c}^{0}}/dy}{d\sigma_{D_{s}^{+}}/dy}=\frac{\lambda_{s}}{2+\lambda_{s}}R_{B/M}^{(c)},
\end{equation}
this ratio is smaller than $\Xi_{c}^{0}/D_{s}^{+}$ by factor $\lambda_{s}/2$.
In addition, we see that the peak position of ratio $\Omega_{c}^{0}/D_{s}^{+}$
is located at about $p_{T}\approx$ 3-4 GeV/c, which is larger than
the peak position of ratios $\Lambda_{c}^{+}/D^{0}$ and $\Lambda_{c}^{+}/\left(D^{0}+D^{+}\right)$
by about 1 GeV/c. This is the kinetic effect caused by the difference
between up quark distribution and strange quark distribution shown
in Fig. \ref{fig:fu_fs}. 

\section{Summary \label{sec:summary}}

In this paper, we have applied a quark combination model with equal-velocity
combination approximation to study the production of light-flavor
hadrons and single-charmed hadrons in $pp$ collisions at $\sqrt{s}=5.02$
TeV. The systematic comparison with available experimental data indicates
the effectiveness of the model, which is consistent with our previous
studies in $pp$ collisions at $\sqrt{s}=$ 7 and 13 TeV \citep{Li:2017zuj,Gou:2017foe,Song:2018tpv,Zhang:2018vyr}. 

By examining the preliminary data for $p_{T}$ spectra of $\Omega$
and $\phi$ at midrapidity, we found that two spectra exhibit a quark
number scaling property, which gives a first signal for the quark
combination mechanism in $pp$ collisions at $\sqrt{s}=5.02$ TeV.
This scaling property further enables us to conveniently extract the
$p_{T}$ spectrum of strange quarks at hadronization. By fitting experimental
data of hadrons containing up/down quarks such as proton, we also
obtained $p_{T}$ spectrum of up/down quarks. Using the extracted
spectra of up/down and strange quarks, we calculated $p_{T}$ spectra
of $K^{*0}$, $\Lambda$ and $\Xi$ which contain both up/down quarks
and strange quarks and we found a good agreement with their experimental
data. We studied the $p_{T}$ dependence of the $\Omega/\phi$ ratio
and found that the increase/decrease behavior of the ratio with $p_{T}$
is closely related to the concave/convex shape of the logarithm of
strange quark distribution. We also studied the difference between
$p_{T}$ spectrum of up/down quarks and that of strange quarks and
used it to explain the difference among $p_{T}$ spectra of different
kinds of baryons. 

Using the EVC model, we extracted differential cross-section of charm
quarks as the function of $p_{T}$ by fitting experimental data of
$D^{*+}$. We found it is quite consistent in shape with calculations
of perturbative QCD method FONLL. Applying the equal-velocity combination
of charm quarks and light-flavor quarks, we successfully explained
the experimental data for differential cross-sections of $D^{0,+}$,
$D_{s}^{+}$ and $\Lambda_{c}^{+}$ as the function of $p_{T}$. We
predicted differential cross-sections of $\Xi_{c}^{0,+}$ and $\Omega_{c}^{0}$.
Compared with $\Lambda_{c}^{+}$, production of $\Xi_{c}^{0,+}$ and
$\Omega_{c}^{0}$ is suppressed because the abundance of strange quarks
at hadronization is suppressed compared with up/down quarks. We predicted
ratio $\Xi_{c}^{0,+}/D^{0}$ is about 0.16 and $\Omega_{c}^{0}/D^{0}$
is about 0.015 due to the cascade suppress of strangeness. We also
proposed several ratios such as $\Xi_{c}^{0}/D_{s}^{+}$, $\Omega_{c}^{0}/D_{s}^{+}$
to further show the effect of cascade suppress of strangeness caused
by the number of strange quarks involving combination with charm quarks.
These predictions can be tested by future experimental data at LHC. 

\section{Acknowledgments}

This work is supported in part by Shandong Province Natural Science
Foundation under Grants No. ZR2019YQ06 and and No. ZR2019MA053, the
National Natural Science Foundation of China under Grant No. 11975011
and No. 11805082, and Higher Educational Youth Innovation Science
and Technology Program of Shandong Province (Grants No. 2019KJJ010).

\bibliographystyle{apsrev4-1}
\bibliography{ref}

\end{document}